\documentclass[a4paper,11pt]{article}
\usepackage{pos}

\newcommand{\tr}{{\rm tr}\,}
\newcommand{\la}{\langle}
\newcommand{\ra}{\rangle}

\newcommand{\f}[2]{\frac{#1}{#2}}

\title{Localization of Dirac modes in a finite temperature
  $\mathrm{SU}(2)$ Higgs model} \ShortTitle{Localization of Dirac modes in a finite temperature
  $\mathrm{SU}(2)$ Higgs model}

\author*{Gy\"orgy Baranka}
\author{Matteo Giordano}
\affiliation{Institute of Physics and Astronomy, ELTE E\"otv\"os
  Lor\'and University,\\ P\'azm\'any P\'eter s\'et\'any 1/A, H-1117,
  Budapest, Hungary}

\emailAdd{baranka@caesar.elte.hu}
\emailAdd{giordano@bodri.elte.hu}

\abstract{Low-lying Dirac modes become localized at the
  finite-temperature transition in QCD and other gauge theories,
  indicating a strong connection between localization and
  deconfinement. This phenomenon can be understood through the
  ``sea/islands'' picture: in the deconfined phase, modes become
  trapped on ``islands'' of Polyakov loop fluctuations within a
  ``sea'' of ordered Polyakov loops. To test the universality of the
  ``sea/islands'' mechanism, we investigate whether changes in the
  localization properties of low modes occur across other thermal
  transitions where the Polyakov loop becomes ordered, beyond the
  usual deconfinement transition. The fixed-length SU(2) Higgs model
  is appropriate for this study. After mapping out the phase diagram,
  we find that low Dirac modes become localized in the deconfined and
  Higgs phases, where the Polyakov loop is ordered. However,
  localization is absent in the confined phase. These findings confirm
  the ``sea/islands'' picture of localization.}

\FullConference{The 41st International Symposium on Lattice Field Theory (LATTICE2024)\\
  28 July - 3 August 2024\\
  Liverpool, UK\\}

\begin{document}
\maketitle

\section{Introduction}
\label{sec:intro}

The connection between deconfinement and chiral symmetry restoration
at the finite-temperature transition in QCD~\cite{Aoki:2006we,
  Bazavov:2011nk} is still not fully understood. The behavior of the
low Dirac modes across the transition could be key in understanding
this connection. In fact, chiral symmetry breaking is controlled in
the chiral limit by the behavior near zero of the (normalized)
spectral density $\rho(\lambda)$ of Dirac modes,
\begin{equation}
  \label{eq:1}
  \rho(\lambda) = \lim_{V\to\infty}\f{T}{V}
  \left\la \textstyle\sum_n \delta(\lambda-\lambda_n)\right\ra\,, 
\end{equation}
where $T$ and $V$ are the temperature and the volume of the system,
respectively, as shown by the Banks-Casher relation
$|\la\bar{\psi}\psi\ra| \stackrel{m\to 0}{=} \pi
\rho(0^+)$~\cite{Banks:1979yr} for the chiral condensate,
$\la \bar{\psi}\psi\ra$. While chiral symmetry is only approximate for
physical quark masses, the effects of its spontaneous breaking in the
chiral limit are still visible below the crossover temperature, in
particular through a sizeable density of near-zero modes, while its
restoration in the high-temperature phase leads to the near-zero
spectral region getting depleted. On the other hand, deconfinement
affects the localization properties of the low modes: a large amount
of evidence shows that in gauge theories with a genuine phase
transition, low modes are delocalized in the confined phase and
localized in the deconfined phase, up to a critical point in the
spectrum, $\lambda_c$, known as ``mobility edge'', that appears
precisely at the critical temperature~\cite{Kovacs:2009zj,
  Kovacs:2010wx,Giordano:2016nuu,Kovacs:2017uiz,Giordano:2019pvc,
  Vig:2020pgq,Bonati:2020lal,Baranka:2021san,Cardinali:2021fpu,
  Baranka:2022dib} (see Ref.~\cite{Giordano:2021qav} for a
review). Also in QCD low modes are delocalized in the low-temperature
phase, and become localized in the high-temperature phase, and even
though in this case there is no sharply defined critical temperature,
the appearance of the mobility edge is well within the crossover
region~\cite{GarciaGarcia:2005vj,GarciaGarcia:2006gr,Kovacs:2012zq,
  Giordano:2013taa,Nishigaki:2013uya,Ujfalusi:2015nha,Cossu:2016scb,
  Holicki:2018sms,Kehr:2023wrs,Bonanno:2023mzj}. Being sensitive to
both chiral symmetry restoration and deconfinement, low Dirac modes
could help in unveiling the details of the relationship between the
two phenomena.

The connection between localization and deconfinement can be
understood qualitatively through the ``sea/islands'' picture of
localization~\cite{Bruckmann:2011cc,Giordano:2015vla,Giordano:2016cjs,
  Giordano:2016vhx,Giordano:2021qav,Baranka:2022dib}. In its original
version~\cite{Bruckmann:2011cc}, further developed in Refs.~\cite{
  Giordano:2015vla,Giordano:2016cjs,Giordano:2016vhx,Giordano:2021qav},
this picture explained localization of the low modes in
high-temperature QCD in terms of ``islands'' of Polyakov-loop
fluctuations in the ``sea'' of ordered Polyakov loops, taking values
near 1, that characterizes the deconfined phase. These islands are
``energetically'' favorable for the eigenmodes, supporting eigenvalues
below the Matsubara frequency, and are expected to ``trap'' the
corresponding eigenvectors. A refined version of this picture was
formulated in Ref.~\cite{Baranka:2022dib}, where the role of islands
is played more generally by fluctuations that decrease the correlation
of gauge fields in the temporal direction, including but not limited
to Polyakov-loop fluctuations. This picture applies as well in the
quenched case, where at high temperature an exact center symmetry
breaks down spontaneously, in the physical sector selected by static
fermions; and more generally as long as a strong ordering of the
Polyakov loop near 1 takes place. This leads one to expect
localization of low Dirac modes in the deconfined phase of a generic
gauge theory, an expectation confirmed in all the models investigated
so far, both with and without dynamical fermionic
matter~\cite{Kovacs:2009zj,Kovacs:2010wx,Giordano:2016nuu,Kovacs:2017uiz,
  Giordano:2019pvc,Vig:2020pgq,Bonati:2020lal,Baranka:2021san,
  Cardinali:2021fpu,Baranka:2022dib,Giordano:2021qav,
  GarciaGarcia:2005vj,GarciaGarcia:2006gr,Kovacs:2012zq,
  Giordano:2013taa,Nishigaki:2013uya,Ujfalusi:2015nha,Cossu:2016scb,
  Holicki:2018sms,Kehr:2023wrs,Bonanno:2023mzj}.

Further possible tests of this picture include looking at phases of a
gauge theory where the Polyakov loop gets ordered, but different from
the usual deconfined phase; and looking at theories with a different
matter content. One model that allows one to carry out both tests at
once is the fixed-length $\mathrm{SU}(2)$ Higgs
model~\cite{Fradkin:1978dv}.

\begin{figure}[t]
  \centering
  \includegraphics[width=0.31\textwidth]{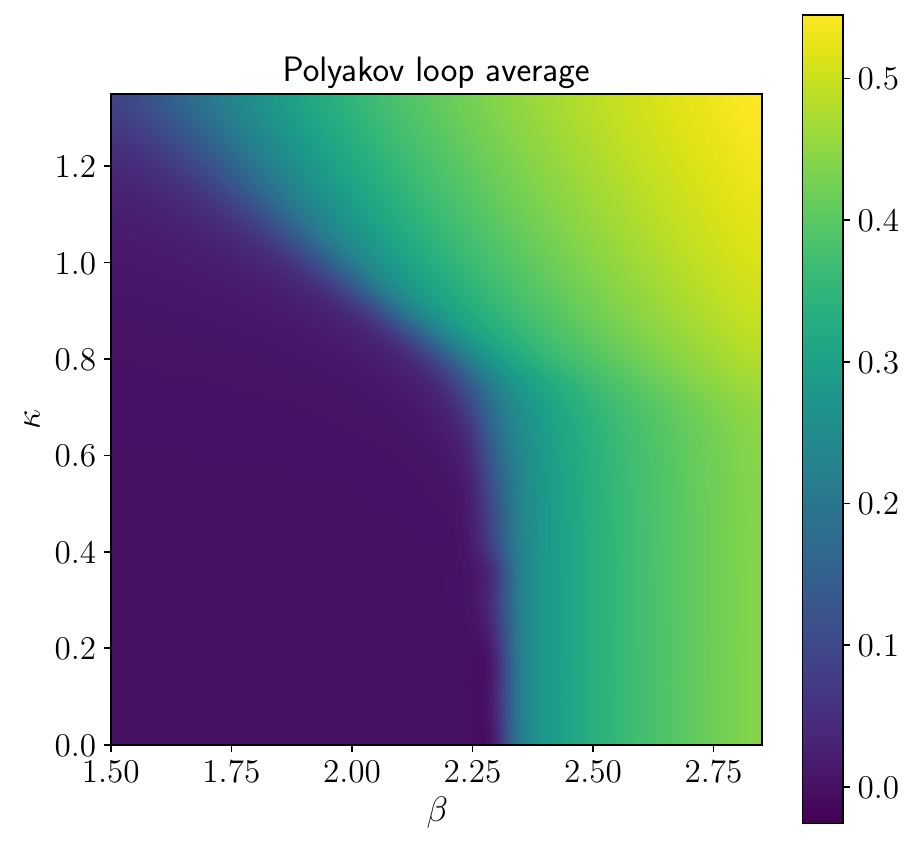}
  \hfil
    \raisebox{-1.5pt}{\includegraphics[width=0.32\textwidth]{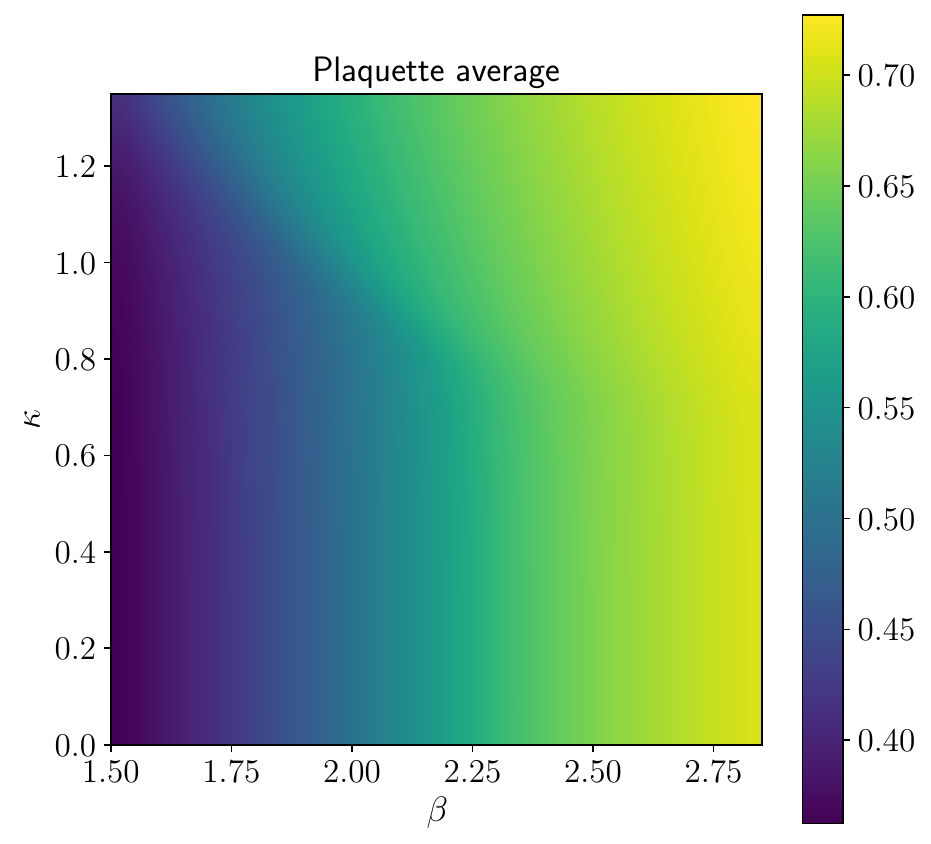}}
  \hfil
  \raisebox{-5pt}{\includegraphics[width=0.32\textwidth]{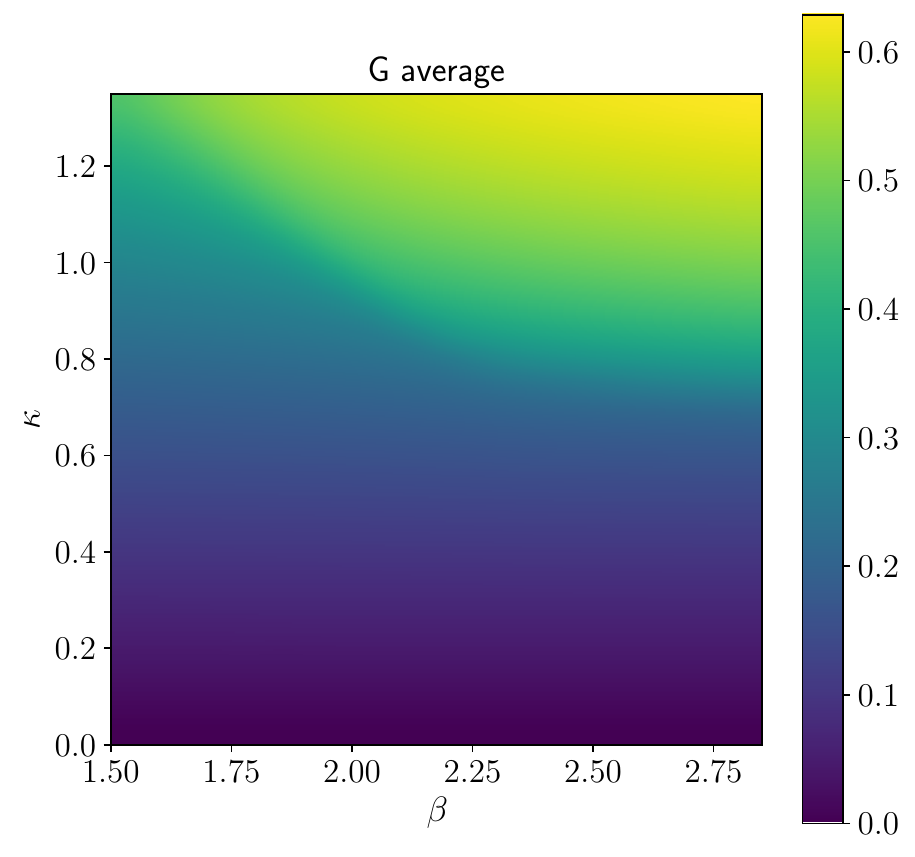}}

  \caption{Expectation values of the Polyakov loop (left), plaquette
    (center), and gauge-Higgs coupling (right) in the $(\beta,\kappa)$
    plane. Here $N_s=20$ and $N_t=4$.}
  \label{fig:1}
\end{figure}

\section{Fixed-length $\mathrm{SU}(2)$ Higgs model}
\label{sec:SU2H}

In Ref.~\cite{Baranka:2023ani} we studied Dirac-mode localization and
tested the sea/islands picture in the fixed-length $\mathrm{SU}(2)$
Higgs model~\cite{Fradkin:1978dv}. This model is obtained from the
usual $\mathrm{SU}(2)$ Higgs model in the limit of infinite Higgs
self-coupling, where the magnitude of the doublet of complex Higgs
fields does not fluctuate. In this case the properly rescaled Higgs
fields can be reorganized in a $\mathrm{SU}(2)$ matrix,
$\phi$. Discretized on a hypercubic $N_t\times N_s^3$ lattice, this
model is defined by the action
\begin{equation}
  \label{eq:SU2Hact}
  S = -\f{1}{2}\sum_n \left\{ \beta\sum_{1\le \mu<\nu\le 4}\tr U_{\mu\nu}(n) + \kappa \sum_{1\le \mu\le 4}\tr G_\mu(n)\right\}\,,
\end{equation}
where $n$ runs over the lattice sites,
$U_{\mu\nu}=U_\mu(n)U_\nu(n+\hat{\mu})U_\mu(n+\hat{\nu})^\dag
U_\nu(n)^\dag$ are the usual plaquette variables built out of the link
variables $U_\mu(n)$,
$G_\mu(n) = \phi(n)^\dag U_\mu(n) \phi(n+\hat{\mu}) $ couples scalar
and gauge fields, and $\beta$ and $\kappa$ are real parameters.

\begin{figure}[t]
  \centering
  \includegraphics[width=0.6\textwidth]{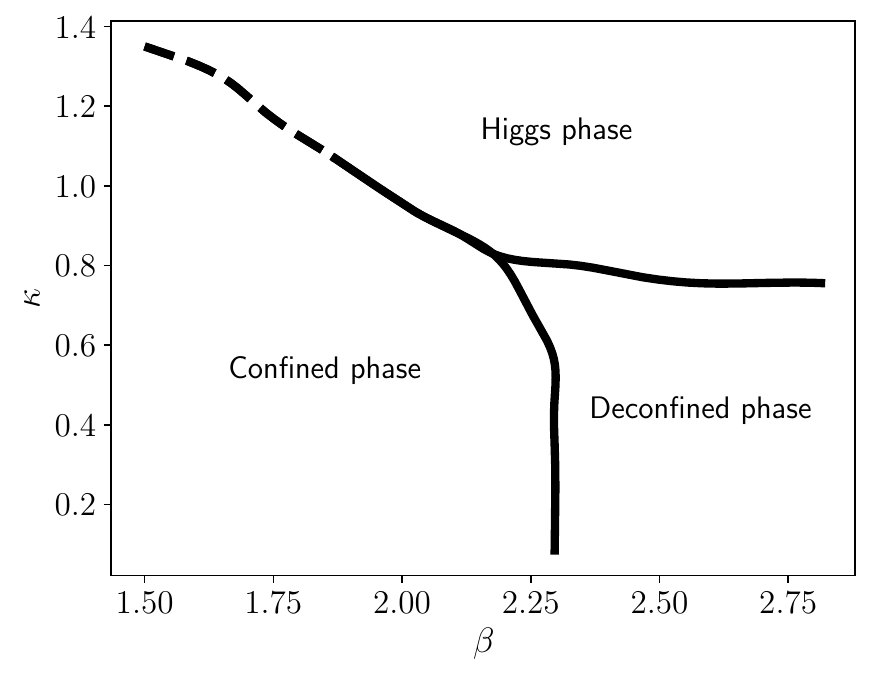}
  
  \caption{Phase diagram of the fixed-length $\mathrm{SU}(2)$
    Higgs model on the lattice for $N_t=4$.}
  \label{fig:2}
\end{figure}

The phase diagram of this model was numerically studied in detail at
zero temperature in Ref.~\cite{Bonati:2009pf}, identifying three
phases -- a confined phase at low $\beta$ and $\kappa$; a deconfined
phase at large $\beta$ and low $\kappa$; and a Higgs phase at large
$\kappa$. This is in agreement with the theoretical expectations of
Ref.~\cite{Fradkin:1978dv}. The phase diagram was studied also at low
temperature resulting in a similar picture, with the transitions
between the three phases identified as crossovers. As a preliminary
step for our study of localization, we mapped out the
finite-temperature phase diagram at fixed temporal extension $N_t=4$
in lattice units. Our results for the average Polyakov loop,
plaquette, and gauge-Higgs coupling $\la G(n) \ra$, where
\begin{equation}
  \label{sec:SU2H2}
  G(n)= \f{1}{8}\sum_{1\le \mu\le 4}\left( G_\mu(n) + G_\mu(n-\hat{\mu})\right)\,,
\end{equation}
are shown in Fig.~\ref{fig:1}. This confirms our expectations for the
phase diagram. We also confirmed the crossover nature of the
transition along three lines, one at constant $\kappa=0.5$ and two at
constant $\beta=2.0$ and $\beta=2.6$, crossing over between the
confined and the deconfined phase, the confined and the Higgs phase,
and the deconfined and the Higgs phase, respectively. Interpolating
the peaks of the Polyakov loop, plaquette, and gauge-Higgs coupling
susceptibilities with splines, and putting the resulting curves
together, we obtained the phase diagram shown in Fig.~\ref{fig:2}.

\section{Localization in the $\mathrm{SU}(2)$ Higgs model}
\label{sec:loc}

To investigate the connection between Polyakov-loop ordering and
Dirac-mode localization we probed the gauge configurations of the
fixed-length $\mathrm{SU}(2)$ Higgs model using external staggered
fermions.  After generating gauge configurations with a standard
heat-bath algorithm, we obtained the low-lying staggered modes with
the PRIMME package for sparse matrices~\cite{PRIMME}, and analyzed
their localization properties as well as the statistical properties of
the spectrum.

\begin{figure}[t]
  \centering
  \includegraphics[width=0.33\textwidth]{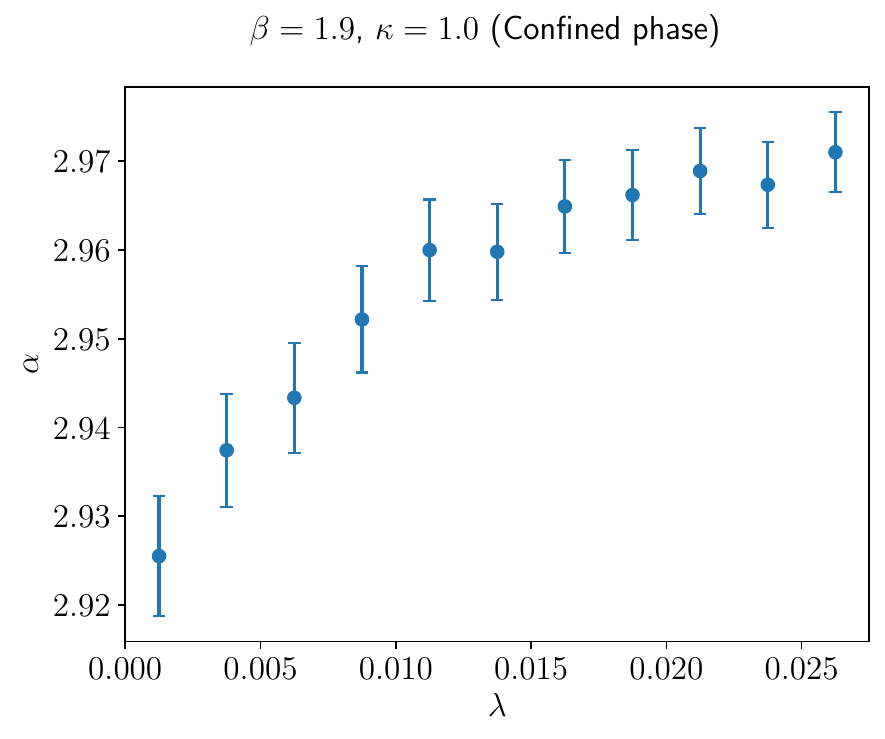}\hfil
  \includegraphics[width=0.33\textwidth]{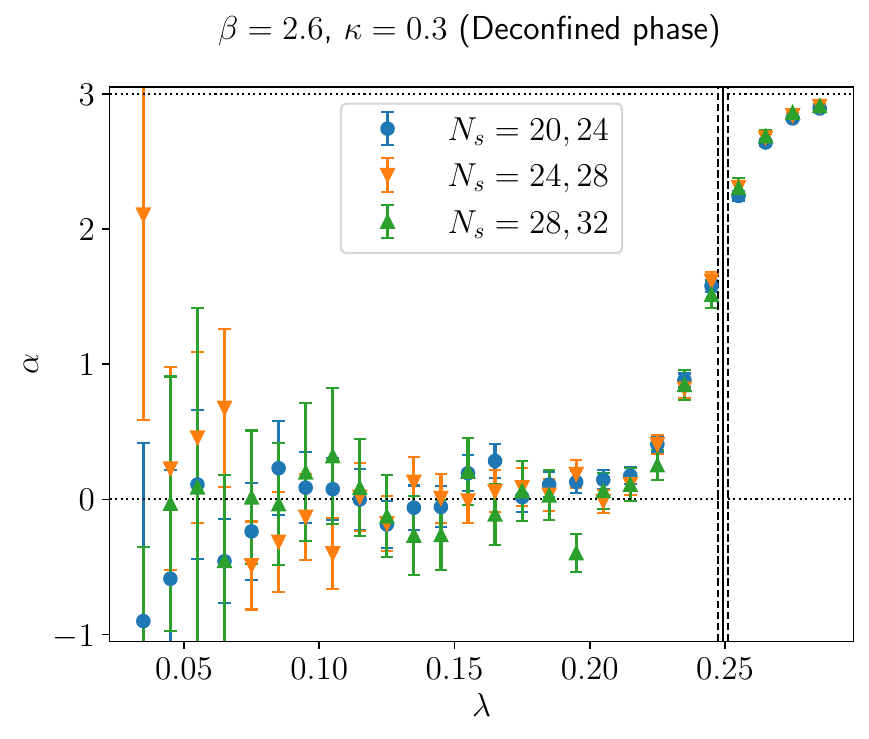}\hfil
  \includegraphics[width=0.33\textwidth]{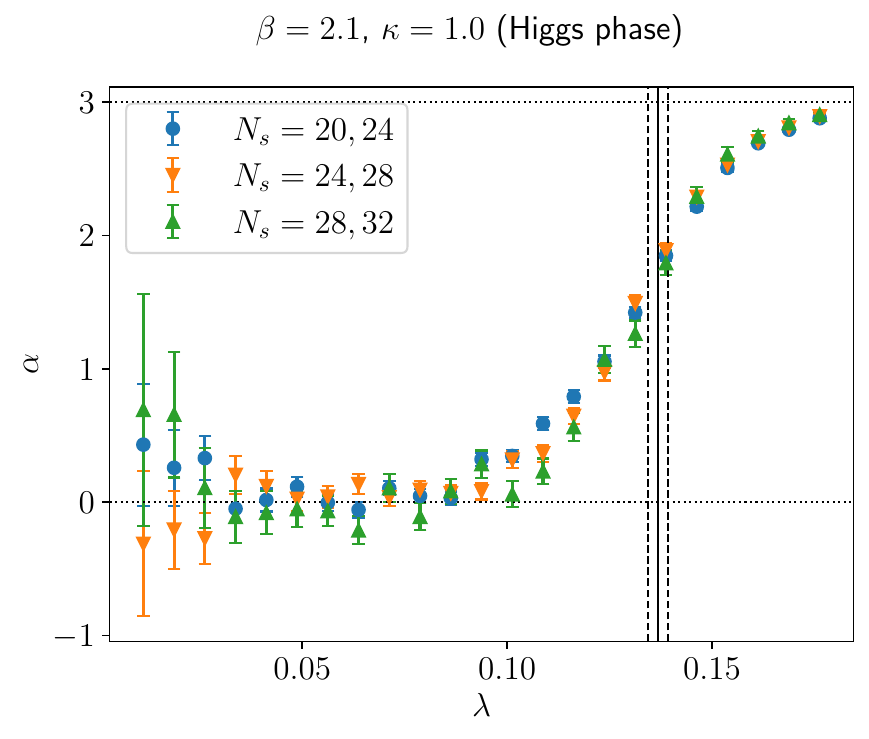}
  \caption{Fractal dimension, Eq.~\eqref{eq:IPR3}, of low staggered
    modes in the confined (left), deconfined (center), and Higgs phase
    (right). Vertical solid and dashed lines mark the position of the
    mobility edge and the corresponding numerical uncertainty,
    respectively.}
  \label{fig:3}
\end{figure}

The localization properties of staggered eigenmodes are defined by the
scaling with the lattice size of the lattice volume that they
effectively occupy. The size of mode $\psi_l$, corresponding to the
staggered eigenvalue $i\lambda_l$, is defined using its inverse
participation ratio,
\begin{equation}
  \label{eq:IPR}
  \mathrm{IPR}_l \equiv \textstyle\sum_{n}\left( \textstyle\sum_c|\psi_{l\,c}(n)|^2\right)^2\,,
\end{equation}
where $c$ is the color index, as
\begin{equation}
  \label{eq:IPR2}
\mathrm{size}_l \equiv \mathrm{IPR}_l^{-1}\,.
\end{equation}
After averaging $\mathrm{size}_l$ in a small spectral interval and
over gauge configurations,
\begin{equation}
  \label{eq:sizelambda}
  \la\mathrm{size}(\lambda)\ra\equiv
  \f{  \la {\textstyle\sum_l}\delta(\lambda-\lambda_l)\mathrm{size}_l\ra}
  {\la{\textstyle\sum_l}\delta(\lambda-\lambda_l)\ra}
  \,, 
\end{equation}
one studies its scaling with the linear spatial size $N_s$ of the
lattice. One expects that
$\la\mathrm{size}(\lambda)\ra \propto N_s^{\alpha(\lambda)}$ for
sufficiently large lattice size, where the fractal dimension
$\alpha(\lambda)$ is $\alpha=0$ if in the spectral region near
$\lambda$ the modes are spatially localized, and $\alpha=3$ if they
are delocalized over the whole lattice. A practical way to estimate
$\alpha$ from numerical data is via
\begin{equation}
  \label{eq:IPR3}
  \alpha(\lambda;N_{s_1},N_{s_2}) \equiv \left.
    \ln\left(\f{\la\mathrm{size}(\lambda)\ra|_{N_{s_2}} }{\la\mathrm{size}(\lambda)\ra|_{N_{s_1}}}\right)
    \middle / \ln \left(\f{N_{s_2}}{N_{s_1}}\right)\right.\,, 
\end{equation}
in the limit of large (and different) $N_{s_{1,2}}$. The dependence of
$\alpha(\lambda;N_{s_1},N_{s_2})$ on the position in the spectrum is
shown for the three phases in Fig.~\ref{fig:3}. While in the confined
phase $\alpha\approx 3$ (estimated using relatively small sizes
$N_{s_{1,2}}=16,20$) in the whole near-zero region, in both the
deconfined and the Higgs phase one clearly observes a transition from
$\alpha=0$ near zero to $\alpha=3$ higher up in the spectrum. This
indicates the presence of a mobility edge, $\lambda_c$, between the
two spectral regions, where the localization length of the localized
modes diverges and a second-order transition (``Anderson
transition''~\cite{Evers:2008zz}) takes place in the spectrum.

\section{Localization and spectral statistics}
\label{sec:stat}

An alternative way to detect localization, that also allows for a
precise determination of $\lambda_c$, exploits the connection between
the localization properties of the eigenmodes and the statistical
properties of the corresponding
eigenvalues~\cite{altshuler1986repulsion}. Since a local change in the
gauge field configuration affects a localized mode only if it takes
place within its localization region, eigenvalues corresponding to
different localized modes fluctuate independently, and one expects
them to obey Poisson statistics. On the other hand, every delocalized
mode is affected by any local change in the gauge configuration, and
so one expects delocalized modes to be strongly correlated with each
other, and to obey the type of random matrix theory
(RMT)~\cite{mehta2004random,Verbaarschot:2000dy} statistics
appropriate for the symmetry class of the system.

A convenient way to exploit this connection is to look at universal
statistical properties of the spacing between subsequent eigenvalues,
unveiled by means of the so-called unfolding procedure. For large
volumes the unfolded level spacings read simply
$s_i \approx (\lambda_{i+1}-\lambda_i)\rho(\lambda_i)V/T$, and their
probability distribution $p(s)$ is of the form
$p_{\mathrm{Poisson}}(s)=e^{-s}$ for Poisson statistics, while for RMT
statistics it is closely approximated by the so-called Wigner surmise,
$p_{\mathrm{RMT}}(s) = a_\beta s^{\beta} e^{-b_\beta s^2}$, where the
Dyson index $\beta$ depends on the symmetry class of the
system~\cite{mehta2004random}. For the staggered operator in the
background of $\mathrm{SU}(2)$ gauge field configurations this is the
symplectic class, for which $\beta=4$~\cite{Verbaarschot:2000dy}.  The
coefficients $a_\beta$ and $b_\beta$ are fixed by the normalization
conditions $\int_0^\infty ds\,p(s)=\int_0^\infty ds\,sp(s)=1$.

\begin{figure}[t]
  \centering
  \includegraphics[width=0.3375\textwidth]{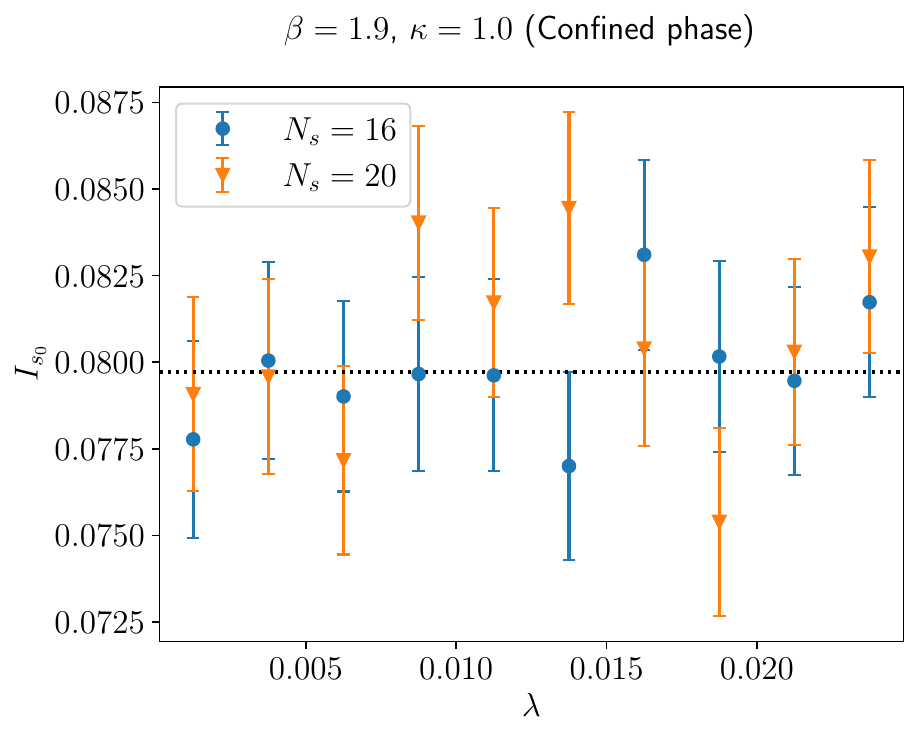}\hfil
    \includegraphics[width=0.33\textwidth]{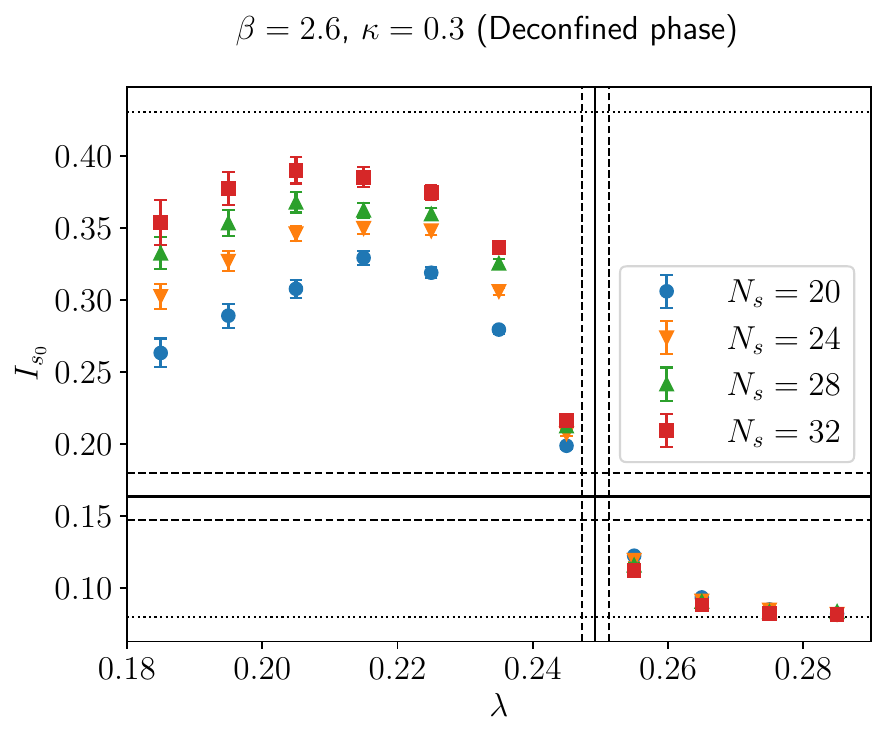}\hfil
    \includegraphics[width=0.33\textwidth]{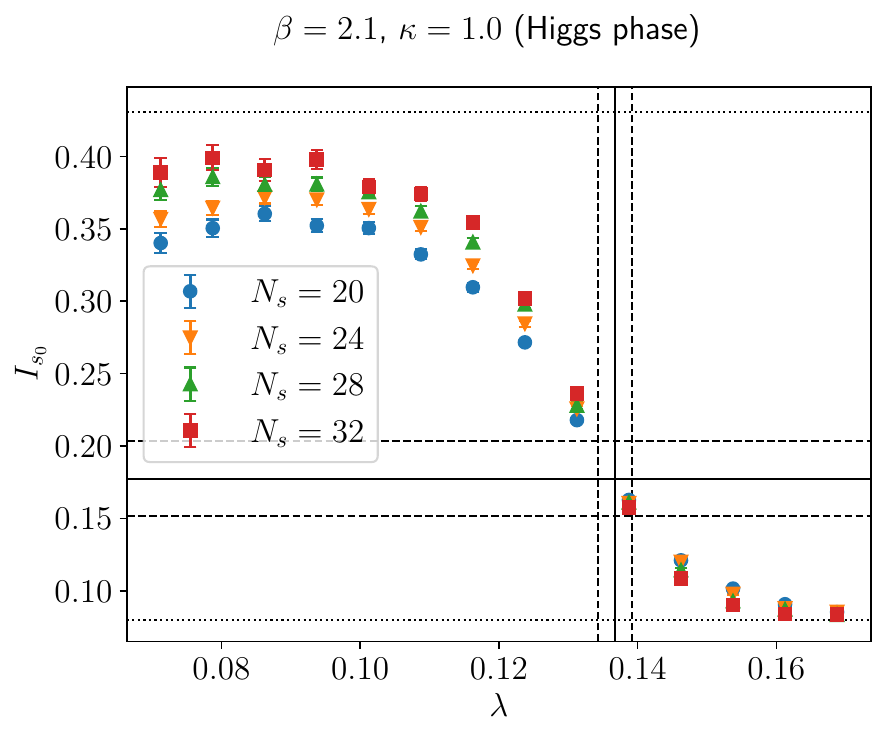}
  
    \caption{Spectral statistic $I_{s_0}$, Eq.~\eqref{eq:Is0}, of low
      staggered modes in the confined (left), deconfined (center), and
      Higgs phase (right). Vertical solid and dashed lines mark the
      position of the mobility edge and the corresponding numerical
      uncertainty, respectively; horizontal solid and dashed lines
      mark the value of the critical value $I_{s_0,c}$ and the
      corresponding numerical uncertainty, respectively.}
  \label{fig:4}
\end{figure}
To identify the mobility edge, one monitors how $p(s)$, evaluated
locally, changes across the spectrum. This is done in practice by
observing some feature of $p(s)$, for example the integrated
probability density $I_{s_0}$~\cite{Shklovskii:1993zz},
\begin{equation}
  \label{eq:Is0}
  I_{s_0} = \int_0^{s_0}ds\,p(s)\,,
\end{equation}
where $s_0$ is conveniently chosen to maximize the difference between
the Poisson and RMT expectations. As one moves from the localized to
the delocalized part of the spectrum, $I_{s_0}$ correspondingly
changes from close to its Poisson value to close to its RMT value,
getting closer and closer as the lattice size is increased. Since a
second-order transition takes place at $\lambda_c$, physics should be
scale-invariant there, and so $I_{s_0}$ at $\lambda_c$ should be
volume-independent. This allows one to accurately identify $\lambda_c$
by means of a finite-size-scaling study~\cite{Shklovskii:1993zz}; or,
less accurately but more practically, by looking at the crossing
points of the curves for $I_{s_0}$ corresponding to different
volumes. We followed the latter approach. In any case, since the
critical statistics is expected to be universal, this kind of study
has to be done only for one set of parameters, and the resulting
critical value $I_{s_0,c}$ of $I_{s_0}$ can then be used to estimate
$\lambda_c$ elsewhere using a single volume by looking for the point
where $I_{s_0}$ crosses $I_{s_0,c}$. For a consistency check, we did
the volume study at two points, one in the deconfined and one in the
Higgs phase.

Our results are shown in Fig.~\ref{fig:4}. In the confined phase
$I_{s_0}\approx I_{s_0,\mathrm{RMT}}$ throughout the low end of the
spectrum, confirming the delocalized nature of these modes. Both in
the deconfined and in the Higgs phase, as the volume increases
$I_{s_0}$ tends to the Poisson value near zero, and to the RMT value
higher up in the spectrum, again confirming our results for the
fractal dimension. The critical values $I_{s_0,c}$ obtained in the two
phases are consistent with each other.

\begin{figure}[t]
  \centering
  \includegraphics[width=0.45\textwidth]{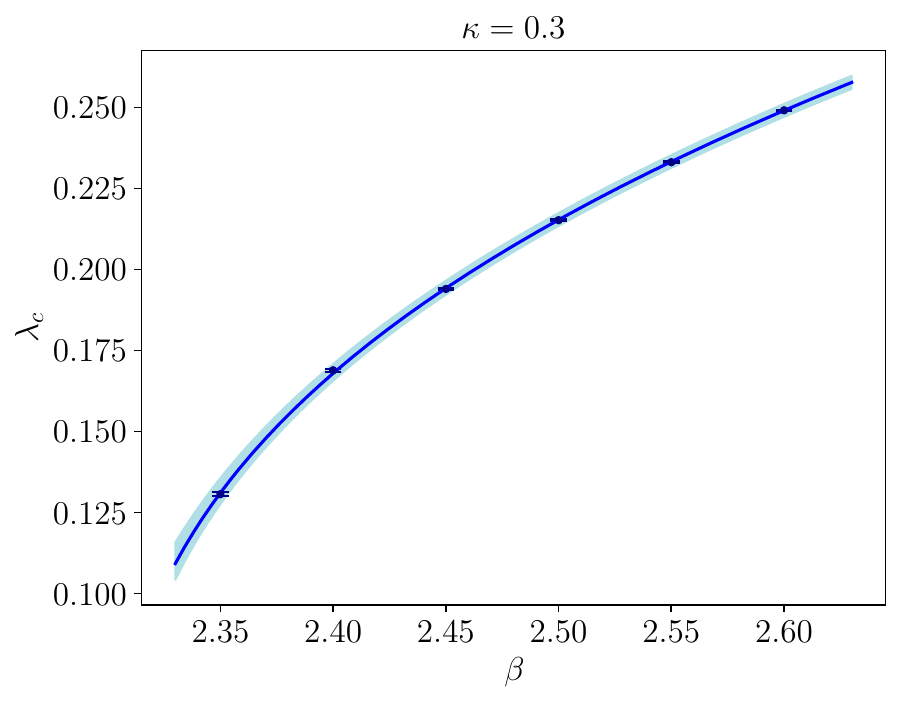}\hfil
\includegraphics[width=0.41\textwidth]{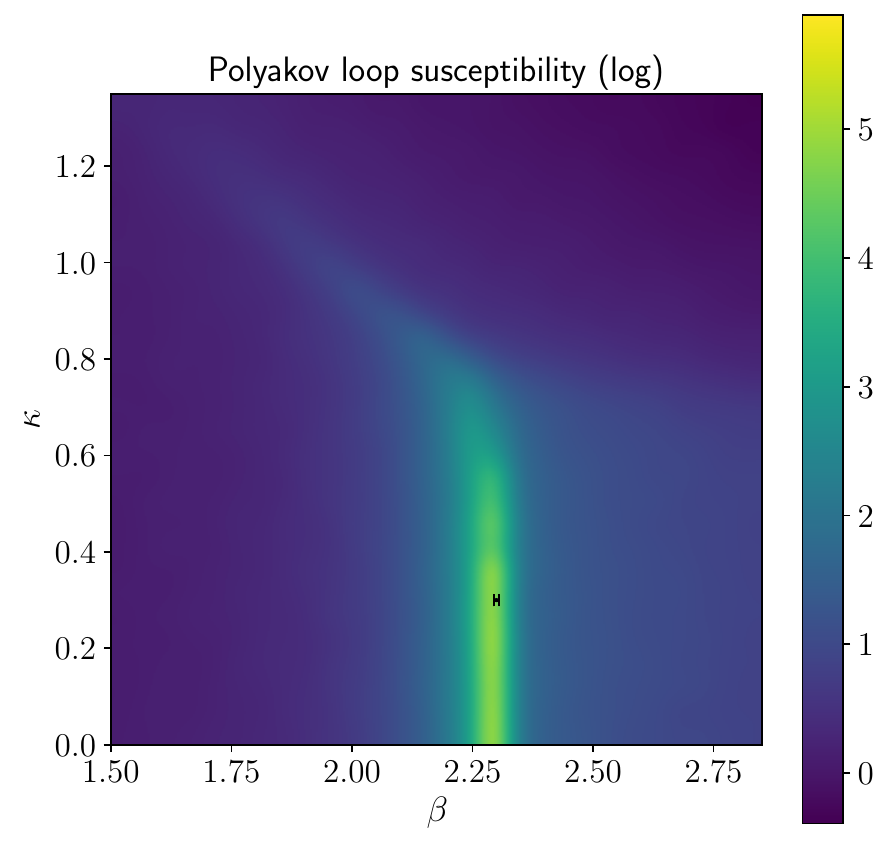}
\caption{$\beta$ dependence of the mobility edge at fixed $\kappa=0.3$
  (left), and Polyakov-loop susceptibility (right).  In the left panel
  we also show a fit with Eq.~\eqref{eq:plaw}. In the right panel we
  show the position of the critical point
  $(\beta_c(\kappa=0.3),\kappa=0.3)$ where the mobility edge
  disappears.}
  \label{fig:5}
\end{figure}

\section{$\beta$ and $\kappa$ dependence of the mobility edge}
\label{sec:bkdep}

According to the sea/islands picture of localization, one expects
$\lambda_c$ to be pushed towards zero as one gets closer to the
confined phase, and to disappear somewhere in the crossover region. As
explained above, one can efficiently estimate the position of the
mobility edge using the critical value $I_{s_0,c}$, and so one can
verify the validity of the sea/islands picture by studying the
dependence of $\lambda_c$ on $\beta$ and $\kappa$ using a single
lattice volume for each choice of parameters. In this study we used a
fixed spatial size $N_s=20$.

The results of our analysis for the transition between the deconfined
and the confined phase at $\kappa=0.3$ are shown in Fig.~\ref{fig:5}
(left panel), together with a plot of the Polyakov loop susceptibility
(right panel). Fitting the numerical data for the mobility edge with a
power law,
\begin{equation}
  \label{eq:plaw}
  \lambda_c(\beta)=a(\beta-\beta_c)^b\,, 
\end{equation}
we find that $\beta_c$, where the mobility edge disappears, is in the
crossover region.

\begin{figure}[t]
  \centering
  \includegraphics[width=0.45\textwidth]{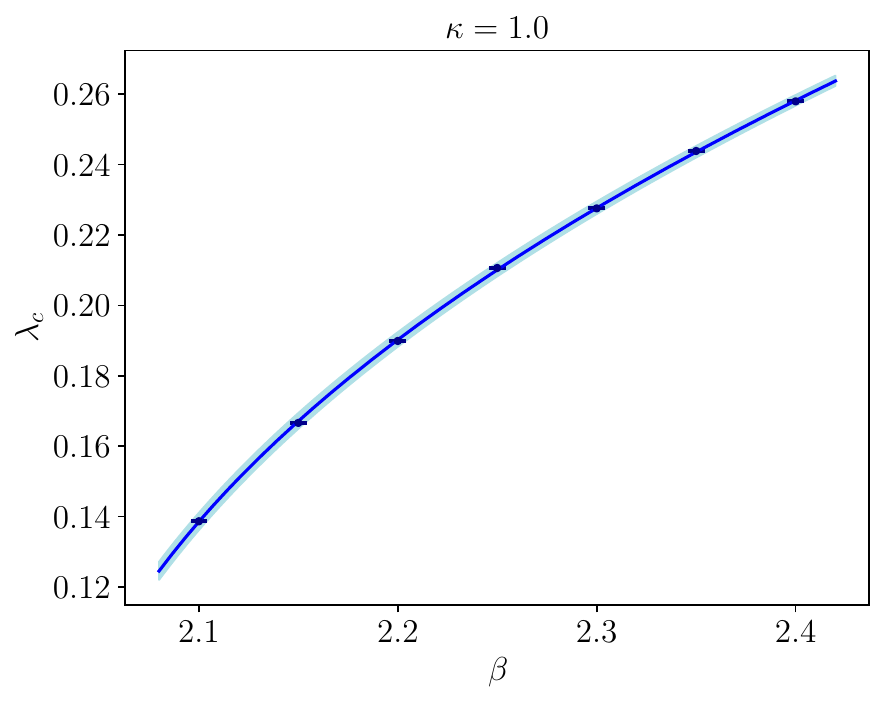}  
  \hfil
  \includegraphics[width=0.44\textwidth]{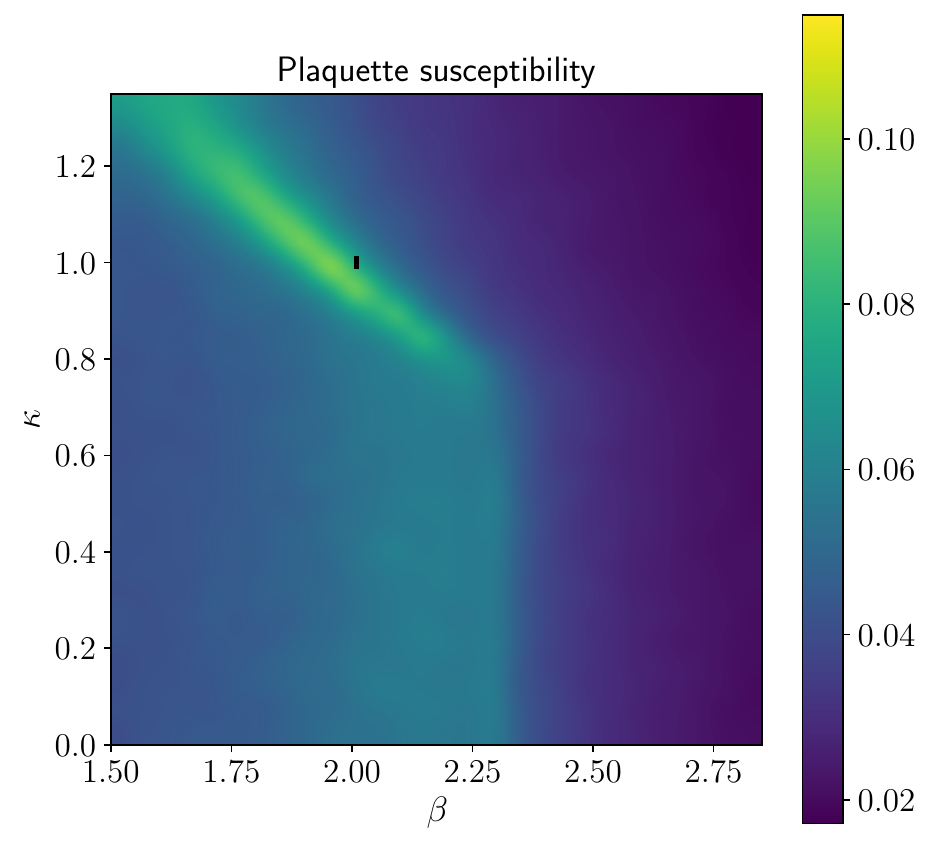}
  \caption{$\beta$ dependence of the mobility edge at fixed
    $\kappa=1.0$ (left), and plaquette susceptibility (right).  In the
    left panel we also show a fit with Eq.~\eqref{eq:plaw}. In the
    right panel we show the position of the critical point
    $(\beta_c(\kappa=1.0),\kappa=1.0)$ where the mobility edge
    disappears.}
  \label{fig:6}
\end{figure}

Our results for the transition between the Higgs and the confined
phase at $\kappa=1.0$ are shown in Fig.~\ref{fig:6} (left panel),
together with a plot of the plaquette susceptibility (right panel). We
fitted our numerical data with the same power-law form
Eq.~\eqref{eq:plaw}, finding again that the point where localized
modes disappear, $\beta_c$, is in the crossover region.

\begin{figure}[t]
  \centering
  \includegraphics[width=0.45\textwidth]{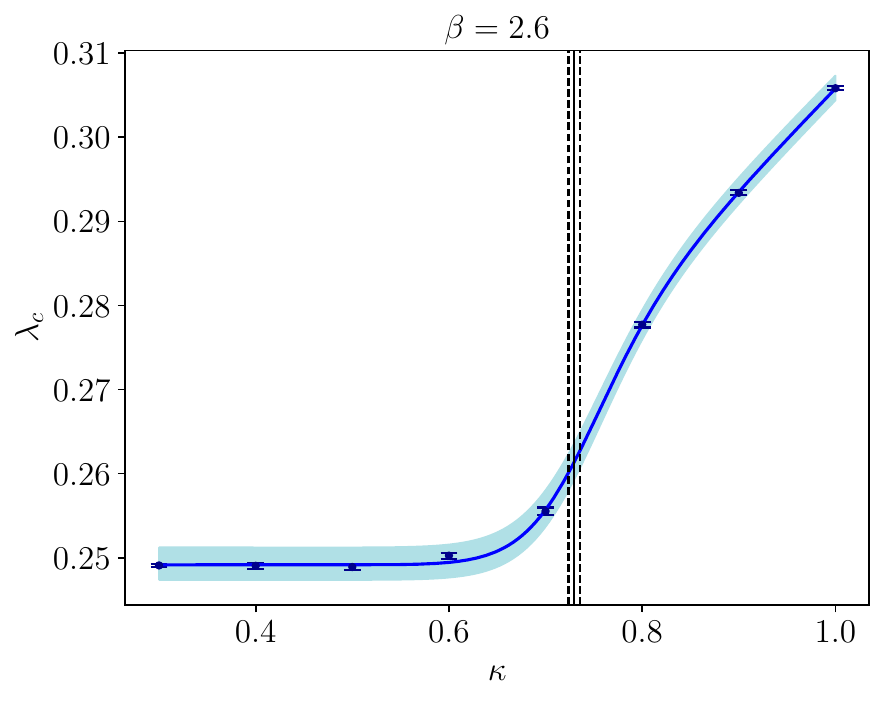}  
  \hfil
  \includegraphics[width=0.43\textwidth]{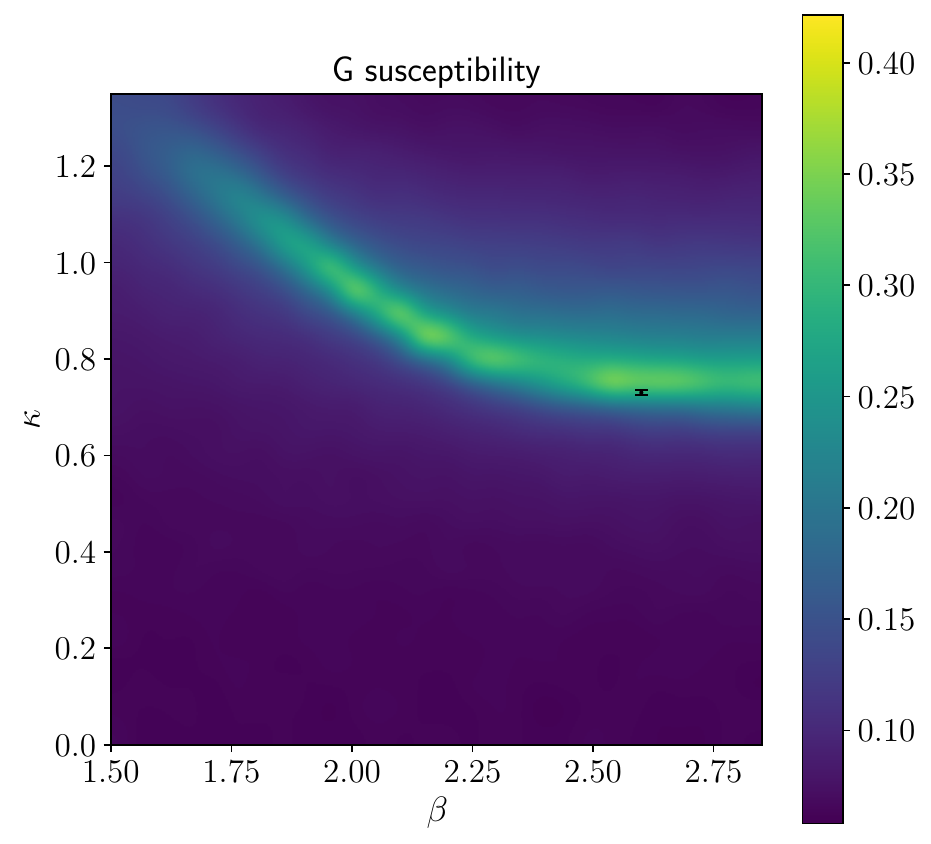}
  
  \caption{$\kappa$ dependence of the mobility edge at fixed
    $\beta=2.6$ (left), and gauge-Higgs coupling susceptibility
    (right). In the left panel we also show the result of a fit with
    Eq.~\eqref{eq:sigmoid} with a solid blue line, and the position
    and uncertainty of the critical value $\kappa_c(\beta=2.6)$, where
    the functional form of the $\kappa$ dependence of the mobility
    edge changes, with vertical solid and dashed black lines,
    respectively. In the right panel we also show the position of the
    critical point $(\beta=2.6,\kappa_c(\beta=2.6))$.}
  \label{fig:7}
\end{figure}

Finally, we show our results for the transition between the deconfined
and the Higgs phase at $\beta=2.6$ in Fig.~\ref{fig:7} (left panel),
together with a plot of the gauge-Higgs coupling susceptibility (right
panel). In this case the mobility edge never vanishes, but the form of
its functional dependence on $\kappa$ clearly changes from almost
constant to almost linear across the transition. Fitting the data with
the function
\begin{equation}
  \label{eq:sigmoid}
  \lambda_c(\kappa) = a\left[1-\sigma\left(d\cdot (\kappa-\kappa_c)\right)\right]
  + (b\kappa+c)\sigma\left(d\cdot (\kappa-\kappa_c)\right)\,,
  \qquad \sigma(x) = \left(1+e^{-x}\right)^{-1}\,,
\end{equation}
we find that the ``critical'' $\kappa_c$ is again in the crossover
region.

\section{Conclusions}
\label{sec:concl}

In this contribution, based on Ref.~\cite{Baranka:2023ani}, we have
once more confirmed the general expectations of the sea/islands
picture of localization, extending the connection between localization
and Polyakov-loop ordering in two directions. In fact, the
fixed-length SU(2) Higgs model investigated here contains scalar
rather than fermionic dynamical matter; and allows one to study a
phase where the Polyakov loop is ordered that is qualitatively
different from the usual deconfined phase. This further confirms that
localization of low Dirac modes is intimately related to the ordering
of the Polyakov loop, without particular regard for the matter content
or other details of the model.

Our study could be extended in two interesting directions. One is to
study the localization properties of the eigenmodes of the covariant
Laplacian at finite temperature, extending the zero-temperature
results of Refs.~\cite{Greensite:2005yu,Greensite:2006ns}, looking in
particular at how these properties change across the various
transitions. Another direction is a detailed study of the low
$\beta$, large $\kappa$ region, where at zero temperature the
transition line obtained with a gauge-invariant spin-glass-inspired
order parameter disagrees with that obtained in a gauge-fixed
setting~\cite{Greensite:2020nhg}, and a study of localization could
shed further light on this discrepancy.

\newpage 
\bibliographystyle{JHEP_nana}
\bibliography{references_LAT24_bar}

\end{document}